\documentclass{article}
\usepackage{graphicx}
\usepackage[utf8]{inputenc}
\usepackage{todonotes}
\usepackage[frozencache,cachedir=minted-cache]{minted}
\usepackage{comment}
\usepackage{url}

\usepackage{amsthm,amsmath,amssymb}
\theoremstyle{plain}
\newtheorem{thm}{Theorem}[section] 
\usepackage{bm}

\theoremstyle{definition}
\newtheorem{defn}[thm]{Definition}

\newcommand{\behavior}{\mintinline{haskell}{Behavior}}
\newcommand{\behaviors}{\mintinline{haskell}{Behavior}s}{}
\newcommand{\hask}[1]{\mintinline{haskell}{#1}}
\newcommand{\sequence}{\texttt{Sequence}}
\newcommand{\fallback}{\texttt{Fallback}}
\newcommand{\rselect}{\mintinline{haskell}{rSelect}}

\newcommand{\seq}{\to}
\newcommand{\fb}{\mathbin{\bm{?}}}

\title{A Behavior Tree-inspired programming language for autonomous agents}
\author{Oliver Biggar and Iman Shames}
\date{}

\begin{document}

\maketitle

\begin{abstract}
    We propose a design for a functional programming language for autonomous agents, built off the ideas and motivations of Behavior Trees (BTs). BTs are a popular model for designing agents behavior in robotics and AI. However, as their growth has increased dramatically, the simple model of BTs has come to be limiting. There is a growing push to increase the functionality of BTs, with the end goal of BTs evolving into a programming language in their own right, centred around the defining BT properties of \emph{modularity} and \emph{reactiveness}. 
    
    In this paper, we examine how the BT model must be extended in order to grow into such a language. We identify some fundamental problems which must be solved: implementing `reactive' selection, 'monitoring' safety-critical conditions, and passing data between actions. We provide a variety of small examples which demonstrate that these problems are complex, and that current BT approaches do not handle them in a manner consistent with modularity. We instead provide a simple set of modular programming primitives for handling these use cases, and show how they can be combined to build complex programs. We present a full specification for our BT-inspired language, and give an implementation in the functional programming language Haskell. Finally, we demonstrate our language by translating a large and complex BT into a simple, unambiguous program.
\end{abstract}

\tableofcontents

\section{Introduction}

How do we express agent behavior in a programming language? Understanding this question is key to designing reliable autonomy in robotics. Compared to traditional programming, autonomous systems present new challenges and requirements~\cite{lee2002embedded}, in particular an emphasis on \emph{real-time behavior}, \emph{safety} and \emph{transparency}. We would like to be able to design our autonomy in a language which is centred on these fundamental principles.

One model for behavioral programming is the Behavior Tree (BT)~\cite{btbook}. This tree-structured design originated in game AI, and has become significantly popular in both game AI and robotics~\cite{iovino2022survey,ogren2022behavior}. Their popularity reflects their simple and effective design philosophy, which places the key principles of \emph{modularity} and real-time \emph{reactiveness} in a central role~\cite{colledanchise2016behavior,biggar2022modularity}. The building blocks of BTs (the leaves of the tree) are asynchronous actions, which can take time and succeed or fail. We build more complex behaviors from these actions through by composition, using a small number of `control flow nodes'. This gives the model a simple tree structure, and reasoning about behavior remains straightforward, even as the tree grows. Further, the two main control flow nodes, Sequence and Fallback, have built-in real-time reactiveness to unexpected events~\cite{ogren2022behavior}, reflecting the importance of this kind of behavior in building autonomous systems.

The importance of modularity and reactiveness is one of the main differences between BTs and the previously-dominant behavioral model, Finite State Machines (FSMs)~\cite{hopcroft2001introduction}. FSMs are an expressive model with uses in many fields, but this flexibility can lead to abuse, and in practice FSMs can grow to great size and complexity as the modelled behavior grows~\cite{btbook}. The structure of FSMs is similar to early imperative languages which made extensive use of the \texttt{goto} statement~\cite{dijkstra1970notes}, resulting in unreadable `spaghetti code'~\cite{btbook}. BTs, by contrast, have a structure which resembles a function call, and this programming analogy is a one of the original motivations for BTs~\cite{ogren2012increasing}. This reflects a general programming principle learned in the structured programming revolution: a language should not only make good programming easy, it should make bad programming difficult. BTs isolated the strengths and weakness of FSMs, and built these into a new model which strictly enforced a modular design.



More generally, and perhaps unsurprisingly, programming language design has been a significant source of inspiration for both BT concepts and philosophy \cite{btbook,ogren2012increasing,biggar2020principled,biggar2022modularity}. A new ecosystem of tools has grown around the simple core model of BTs, augmenting them with constructs for handling data~\cite{shoulson2011parameterizing,schulz2024behavior}, concurrency~\cite{colledanchise2021handling}, parallelism~\cite{improvingparallel}, learning \cite{sprague2022adding,kartavsev2024improving}, formal methods \cite{biggar2020framework,serbinowska2022behaverify}, and much more~\cite{colledanchise2021implementation,iovino2022survey,btbook,ghzouli2020behavior}. Moreover, tools exist to automatically generate code from BTs, written in a visual editor or XML~\cite{btcpp}. The result is that BTs are approaching a programming language in their own right, reflecting the value of the original reactiveness and modularity principles.


The myriad of BT extensions are necessary precisely because BTs are a \emph{model} and not a programming language. This means that their core semantics are comparatively underspecified. This makes sense---BTs are a tool for us to think in a modular way about behavior design, and this is what they achieve. The presence of these extensions, as we see it, is a call for more functionality from BTs, that is, for them to \emph{evolve into a fully-fledged programming language} centred on the principles of behavioral programming.

This evolution into a programming language is already taking place, in an \textit{ad hoc} and piecemeal way~\cite{colledanchise2021implementation}. However, there is a problem. The tree presentation of BTs, useful as a conceptual tool, presents a challenge for extensions. Importing additional functionality---including basic concepts from programming, like loops and variables---requires the overhead of building new control flow nodes. Accordingly, the simple BT model is being increasingly swamped by a proliferation of constructs borrowing different programming concepts and translating them into tree form~\cite{unifiedframework,gugliermo2024evaluating}. By reinventing the wheel, BT extensions risk wasting half a century of insight from programming language design, leading to solutions which are not consistent with good programming practice~\cite{shoulson2011parameterizing,unifiedframework,colledanchise2021implementation,anguelovBTmisuse,biggar2021expressiveness,biggar2020principled}. Examples include \cite{btbook}: Blackboards, the standard mechanism for introducing data to BTs, which is built on \emph{global variables} (Section~\ref{sec: data}); using Success and Failure for both booleans and exceptions, leading to cases where \emph{we can't distinguish between a failed behavior or a false value} (Section~\ref{sec: booleans}); using Sequence and Fallback to implement branching in the absence of an \texttt{if}-statement; using auxiliary variables to implement stateful behavior (Section~\ref{sec: when to use react}); and more. Similarly, the `tick' mechanism used to create reactiveness can be misleading in concurrent implementations, leading to chattering or an inability to make progress (Section~\ref{sec: reactiveness}). This problems are in contrast with the programming analogies which motivated the original popularity of BTs.


In our view, these problems arise from the original BT model having outgrown its roots. There is a natural solution, however. Just as BTs isolated the key strengths and weaknesses of FSMs and built them into a stricter framework for behavior modelling, we believe the time has come for a full programming language which is built on the key principles and use cases of BTs. That is, a language whose fundamental units are asynchronous actions, with a small set of built-in constructs to combine those units into larger units in a modular way. Reactiveness, with its key role in BTs, should equally be a central component of this language. However, unlike BTs, a full programming language must also incorporate data and control flow into its core structure, in a way which doesn't undermine the BT principles. This extends the scope of BT design, without needing to reinvent the wheel.

\begin{figure}
    \centering
    \begin{minted}{haskell}
openPassClose door = open door >> passThrough door >> close door
enterRoom = do doors <- findDoors
               fallback doors (attemptNTimes 3 openPassClose)
monitorBattery = monitor ((<10) <$> batteryLevel) recharge
bt = monitorBattery (enterRoom >> performTask)
    \end{minted}
    \caption{An example of a BT programmed in our language. This desired specification is: the robot must enter a room to perform as task (\texttt{performTask}). Entering the room requires entering through some door (\hask{openPassClose}). There are multiple doors into the room, and the robot must try to open each three times before moving on to the next door (\hask{enterRoom}). Finally, while performing this behavior, the robot must monitor its battery level. Whenever the level drops below 10, the robot must recharge fully before continuing (\hask{monitorBattery}). Note how this complex behavior can be implemented in a precise, modular and safe way.}
    \label{fig:code snippet}
\end{figure}

This is our goal in this paper: to isolate the principles and strengths of BTs and build a programming language for autonomous behavior which has those principles at its heart. We implement this as a library within the functional programming language Haskell. See Figure~\ref{fig:code snippet} for an example implementation of a robotic behavior. Our language is built around a \behavior{} type, which represents an action as an asynchronous function, which can both take arguments and return a value. Consistent with both BT's modular design and Haskell's functional paradigm, all more complex programs are built from function composition on \behaviors{}, using a small collection of operations which include both classical sequencing (\hask{>>=}) and selection (\hask{if-then-else}), as we well as two reactive operations (\rselect{} and \hask{monitor}). Haskell has a strict type system to enforce safe modular composition, similar to the return value interface used by BTs. The resultant programs are simple and modular, compiling a collection of robotic actions into a complex decision-making mechanism. We demonstrate our library with examples throughout, as well as a longer example in Section~\ref{sec: example}.

While BTs are the inspiration for our language, our constructs are \emph{not} direct translations of the BT control flow nodes. That is, our language is not just another implementation of BTs. We believe that some specific constructs of BTs either cause problems for concurrency (e.g. the tick mechanism, see Sections~\ref{sec:bts} and \ref{sec: reactiveness}) or become unnecessary in the context of a full programming language (e.g. the Success, Failure and Running flags, see Section~\ref{sec: booleans}). For instance, the standard reactive Sequence and Fallback control flow nodes \emph{are not present} in our language, though we argue that their use cases are still captured by other, more intuitive operations. The main challenge and contribution of this paper is these design decisions, for which we must analyse BT use cases in practice and balance the competing design requirements of simplicity, modularity and ease of use. We particularly focus on techniques for (1) reactiveness (Section~\ref{sec: reactiveness}), a key BT feature which by its concurrent nature can be confusing; and (2) developing a modular composition method for data (Section~\ref{sec: data}), an important programming language concept which is typically a secondary consideration in BTs. By its nature, this language design task is inherently subjective, and some of our conclusions are at odds with previous approaches to BT design. We motivate all of our design decisions with examples and code snippets, so that readers can judge for themselves whether our approach to behavioral programming meets their needs.


One such design decision was the choice of Haskell as an implementation language, which we will briefly discuss here. Haskell is an unusual choice in robotics, where C++ and Python are typically dominant. However, we have some good reasons for this decision. The first is that, as autonomous agents become more complex, behavior design becomes a distinctly different task from action implementation, requiring a different skill-set and performed by different people. Conceptually separating this low- and high-level reasoning is a goal of BTs~\cite{btbook}, and modern BT implementations often already separate BT design from action programming, using a visual editor or, in BT.CPP \cite{btcpp}, a dedicated XML sublanguage. Haskell, with its simple and flexible syntax, has been used before for constructing domain-specific languages in industry~\cite{marlow2010haskell}. Secondly, and more importantly, the design philosophy of BTs, and particularly its emphasis on modularity and composition, is consistent with Haskell's purely functional programming paradigm. The philosophy of functional programming is that we build programs declaratively, by composing simple functions in a modular way, leading to programs which are trees of expressions, like BTs. Compared to imperative programming, functional programming emphasises safety and composability, particularly by avoiding unintended side effects. These properties are especially important in robotics. Most specifically, our main sequencing operation, which combines the Sequence-with-memory control flow node with function composition (Section~\ref{sec: data}) is identical to Haskell's \hask{bind} function, the core of its flagship monadic approach to purely functional I/O.


That said, the main goal of this paper is to isolate the principles and concepts necessary to build a BT-based programming language. Though we think that Haskell is a good choice for implementing these constructs, outlined in Section~\ref{sec: specification}, they could also be used as the core of a behavioral programming library in any other language.

\section{What are Behavior Trees?} \label{sec:bts}

We must firstly describe what a Behavior Tree is. There exist many variations in terminology, but for the purpose here we will follow the standard setup from \cite{btbook}, a well-known textbook on Behavior Trees in robotics. See Figure~\ref{fig:btbooktree} for an example from \cite{btbook}. For the unfamiliar, \cite{btbook} provides a very thorough introduction to BTs and their motivations.

Behavior Trees are a tree-structured \emph{control architecture}, designed to model the discrete task-switching behaviors of autonomous agents. This tree structure provides modularity---each subtree can be treated as an individual behavior, with the internal nodes representing \emph{composition} of behaviors. Execution of a BT proceeds through `ticks', which are generated at the root and repeatedly poll the subtrees\footnote{In Section~\ref{sec: reactiveness} we discuss the challenges associated with the `tick-based' execution model.}. Nodes execute when they are ticked. Internal nodes (called \emph{control flow nodes}) tick their children when ticked, and leaf nodes are called \emph{execution nodes} which perform actions or check conditions. All nodes have a common interface which returns execution values to its parent node. There are three values: Success, indicating the node has completed successfully; Failure, indicating some error; and Running, indicating that execution of the node is still underway. The third is important; real-world tasks can take time, and the Running value allows the parent node to know that the task is proceeding. Each tick is independent, so at each step the tree is ticked again from the root. This provides reactiveness: if the return value of a behavior changes, the tree can respond immediately, regardless of the previously executed node.

The leaves of the tree (the \emph{execution nodes}) are the basic building blocks of the behavior: Actions and Conditions. Actions represent actions performed by the agent, and Conditions check some property. The difference between them is that Conditions are assumed to not return Running, and instead immediately return either Success or Failure\footnote{Not that the Success and Failure flags are used as booleans. We discuss this in Section~\ref{sec: booleans}.}.

Composition of Actions and conditions into complex behaviors occurs through the internal nodes (\emph{control flow nodes}). Four types are typically used~\cite{btbook}

\begin{itemize}
    \item \textbf{Sequence}: The Sequence node (depicted by $\to$), ticks its children from left to right. If a child returns Running or Failure, the Sequence node returns this value. If a child returns Success, Sequence ticks the next child, and it returns Success if all of its children do so.
    \item \textbf{Fallback}: Fallback (written $\bm{?}$) is the complement of Sequence. It ticks its children from left to right, returning Running or Success if a child does, and returning Failure if all of its children return Failure.
    \item \textbf{Parallel}: The Parallel node (written $\rightrightarrows$) executes its children in parallel. Parallel nodes are typically given a `Success threshold' $M$, where for a Parallel node with $N$ children it returns Success if $M$ children do so, Failure if $N-M+1$ do so, and Running otherwise.
    \item \textbf{Decorator}: Decorators are a broad class of nodes with a single child, which allow for essentially any user-defined policy. Some examples of Decorators implemented in \cite{btcpp} are: Inverter (flip the values of Success and Failure); Repeat (repeat a child $N$ times, while it returns Success), RetryUntilSuccessful (repeat a child $N$ times, while it returns Failure) and RunOnce (execute the child once, and thereafter output its return value without ticking it again).
\end{itemize}

\section{Reactiveness} \label{sec: reactiveness}

Reactiveness, in the sense of being able to respond quickly to real-time stimuli, is critical in robotics~\cite{ogren2022behavior}. BTs embrace the central role of reactiveness, and provide the means to construct reactive behavior using a small collection of modular operators. Certainly, reactiveness should equally be a central concept in our language.

However, (non-reactive) stateful behavior is also important, because many simple behaviors can't be constructed without some concept of state~\cite{biggar2021expressiveness}. Unfortunately, classical approaches for defining state---such as FSMs---do not strictly enforce a modular design. Consequently, they can become hard to read as they grow in size. Combining stateful and non-stateful behavior is therefore a necessary but challenging goal in BT research. In particular, we need to clearly delineate between these cases, to ensure the overall system remains easy to understand.

Unfortunately, existing BTs do not always clearly distinguish between reactive and non-reactive behavior, and the result is systems which are opaque and often ambiguous. To develop the reactive constructs for our language, we need to isolate exactly how reactiveness is used in BTs. In this section we identify two fundamental use cases of reactiveness in BTs: partitioning the state space into different region depending on actions, which we call \emph{reactive selection} and interrupting lower-priority tasks to handle higher-priority tasks, which we call \emph{monitoring}. To design \emph{reactive selection}, we use BTs with a decision-tree-like structure, using nested conditions and actions to modularly separate the space into the different cases. \emph{Monitoring} involves running a low-priority action while continually checking the conditions for a higher-priority action, interrupting the lower-priority action whenever the higher-priority action must be performed. Just as in BTs, more complicated behavior can be built from composition of these two operations. The result is a modular separation of reactive behavior, which we can then freely combine with standard stateful constructs from structured programming, like sequencing, branching and looping. We also identify where explicitly stateful behavior is clearer and more modular than reactive behavior, laying out a collection of design principles for behavioral code.

\subsection{Reactive selection and the `progress problem'} \label{sec: ticking}

The \emph{monitoring} and \emph{reactive selection} use cases we described above certainly seem like plausible uses of reactiveness, but why should they be the \emph{only} uses of reactiveness? Why should we view these operations as the fundamental reactive constructions, rather than, for instance, \sequence{} and \fallback? Our argument is that BTs which are not in particular decision-tree-like form---which defines the \emph{reactive selection} operation---suffer from a fundamental problem caused by the tick mechanism which defines \sequence{} and \fallback. Attempts to circumvent this problem lead to confusing, non-modular and ambiguous BTs which mix reactive and non-reactive behavior.

We begin with a seemingly simple question: what does the \sequence{} operation mean? (Note that \fallback{} is the same as \sequence{} with the role of Success and Failure reversed, so we are simultaneously trying to understand \fallback). Consider a \sequence{} composition \texttt{bhvr1 $\to$ bhvr2} of two Actions \texttt{action1} and \texttt{action2}. The classical tick-based definition is the following: on each tick, the sequence node first ticks \texttt{action1}, which either: (i) returns Failure, in which case the sequence returns Failure; (ii) returns Running, in which case it `\emph{performs a piece of computation}'; or (iii) returns Success, in which case \texttt{action2} is ticked. It is implicit here that the actions can be decomposed into atomic tick-sized chunks. While this assumption was reasonable in the original game AI roots \cite{isla2015handling} of BTs, where the virtual world updates in discrete time, it often doesn't make sense in robotics, because the real world doesn't behave this way. Many robotic actions can't be conceptually decomposed into atomic steps \footnote{As an example, consider a \texttt{Hover} behavior for a drone.}. Worse, this isn't modular---the behavior designer must explicitly decompose the behavior into pieces, or else sacrifice the reactiveness of the tree. Understandably, then, supporting asychronous behavior is a standard feature of modern robotics BT libraries \cite{colledanchise2021implementation}, and increasingly of game AI libraries too. Under a concurrent interpretation of \sequence{}, the first tick starts the behavior and subsequent ticks poll it for its status, until it eventually completes or fails. \textbf{But now we have a problem.}

Let's consider what happens when we execute \texttt{action1 $\to$ action2}. Suppose on the first tick we set \texttt{action1} running. Both behaviors can be anything, so in general \texttt{action1} may take some time---and many ticks---to execute. Some number of ticks with \texttt{action1} returning Running when polled, until at some point it returns Success. On that tick we start \texttt{action2}. However, on the next tick, after negligible time, we tick \texttt{action1} again which either immediately returns Success or we cancel \texttt{action2} and set \texttt{action1} running. The result is that, if \texttt{action1} isn't guaranteed to return Success immediately, \emph{we can never make progress} on \texttt{action2}, because after \texttt{action1} completes we immediately restart it. We will call this the \textbf{progress problem}. This seems to prohibit ever placing two actions in \sequence{}, because we can never make any progress on the second.

Firstly, note that the progress problem is a reactive phenomenon; it doesn't happen in the non-reactive case, \texttt{action1 $\to^*$ action2}, where $\to^*$ represents the \emph{sequence with memory} operation. Here we just do \texttt{action1}, and then we do \texttt{action2}. This is sequencing in the standard programming sense.

Secondly, the problem arises from \emph{mixing specification and implementation}. What we wanted from \sequence{} was the intuitive description: do \texttt{action1} whenever its goal is successfully completed, and do \texttt{action2} otherwise. The tick mechanism is one way of implementing this behavior, used in the original game AI BTs. However, by making ticks part of the \emph{definition} of \sequence{} we obtain problems generalising to truly concurrent Actions.

Thirdly, even in BTs which use reactive \sequence{} and \fallback{}, the progress problem doesn't always arise. With a specific BT structure, we can avoid it. Suppose our BT looks like this: \texttt{(condition ? action1) $\to$ action2}, where \texttt{condition} is a Condition and so is \emph{guaranteed to return Success or Failure in one tick}. Because we always tick \texttt{condition} first, and then do either \texttt{action1} or \texttt{action2}, we have avoided the progress problem. This BT is intended to be read as a reactive version of an \texttt{if} statement\footnote{This BT is not exactly an \texttt{if} statement, because it is not symmetric in its arguments. If \texttt{action1} succeeds, we follow it by doing \texttt{action2}. We define reactive selection to be truly symmetric, with neither child behavior preceding or succeeding the other.}, which we might write in pseudocode as:
\begin{minted}{haskell}
bt = reactively if condition then action1 else action2
\end{minted}
The `\texttt{reactively}' is indicating that we check \texttt{condition} at each tick, and we swap between the two actions whenever \texttt{condition} changes value. We arrive at our first use case for reactiveness: \emph{reactive selection}.
\begin{defn}[Reactive Selection]
\emph{Reactive selection} is an operation which takes three arguments: a condition \texttt{test}, and two behaviors \texttt{left} and \texttt{right}. We will write this as \hask{rSelect test left right}, where the arguments are interpreted like an \hask{if-then-else}. \rselect{} continually runs \texttt{left} whenever \texttt{test} is true and \texttt{right} whenever \texttt{test} is false, switching whenever the value of \texttt{test} changes. It fails when either \texttt{left} or \texttt{right} fails, and succeeds when either runs to completion.
\end{defn}
We call this `reactive selection', because it is a reactive version of an \texttt{if}-statement. We can nest reactive selections recursively. So, for instance, adding another layer gives us the BT 
\begin{equation} \label{nested BT eg}
(\tt{c1} \fb ((\tt{c2} \fb \tt{a1}) \seq \tt{a2})) \seq ((\tt{c3} \fb \tt{a4}) \seq \tt{a5})
\end{equation}
The BT in~\eqref{nested BT eg} is equivalent to 
\begin{minted}{haskell}
bt = rSelect c1 (rSelect c3 a5 a4) (rSelect c2 a2 a1)
\end{minted}
When a BT consists only of nested reactive selections, such as in~\eqref{nested BT eg}, we say it is in \emph{reactive selection form}. In these BTs, all actions are in the right-hand branch of their parent \sequence{} or \fallback{} node. Because all other nodes are conditions, we never have the problem of two actions executing in a \sequence{}.

This example demonstrates an important readability point: the original BT~\eqref{nested BT eg} is highly opaque, and its selection structure is hard to discern. By contrast, the \texttt{if}-statement form is much more readable, and is consistent with programming conventions. Consequently, we believe that BTs in reactive selection form should be explicitly represented in this way, like the above. This is an important point, because reactive selection form is a very common way of structuring BTs in practice. For instance, one common design technique for BTs called \emph{backchaining}, always results in reactive selection BTs~\cite{btbook}.

To recap: some BTs avoid the progress problem by using reactive selection form, and this construction can be represented clearly and modularly by a reactive version of an \texttt{if}-statement. However, not all BTs have this structure. How do other BTs avoid the progress problem? We believe that this problem often goes unnoticed because we \emph{implicitly interpret BTs as if they were in reactive selection form}, even when they are not. The result, as we see it, is that BTs which aren't in reactive selection form are inherently ambiguous.

\begin{figure}
    \centering
    \includegraphics{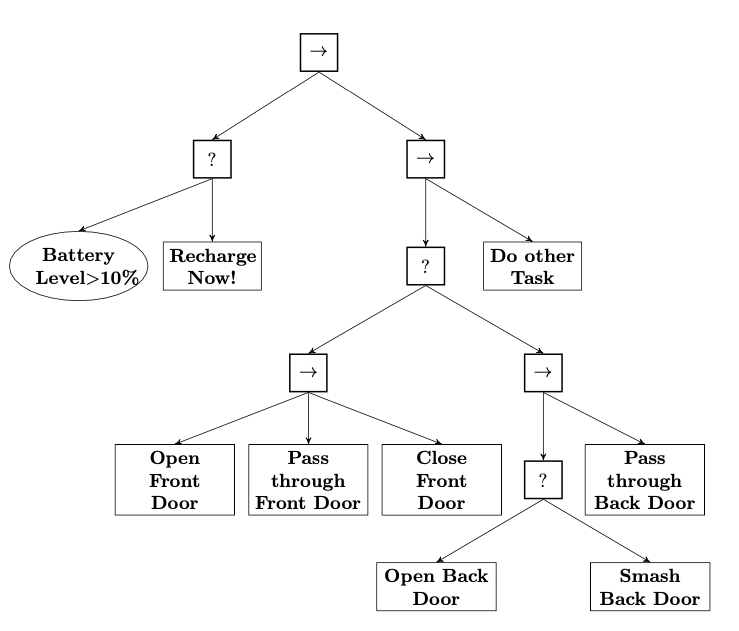}
    \caption{A Behavior Tree for a complex robotic task, from~\cite{btbook}.}
    \label{fig:btbooktree}
\end{figure}
Let's look at a concrete example of a BT, shown in Figure~\ref{fig:btbooktree}, taken from \cite{btbook}. This is a classic BT example: a task is combined from a number of subtasks (opening and passing through a door) as well as managing safety guarantees (the robot must recharge its battery as necessary).


First, observe that the top level of this tree is in reactive selection form: \texttt{(battery>10\% ? Recharge) $\to$ (...)}. However, the subtree responsible for entering the room and performing the task is not. There we find Actions placed in \sequence, each of which may take time to complete. Let's look first at \texttt{open door $\to$ passThrough door}. We open the door, which takes some time, then, once it returns Success, we pass through the door. While passing through door, we still tick \texttt{open door}, and so long as the door remains open we return Success every tick and so are able to make progress on \texttt{passThrough door}. Have we avoided the progress problem? No. Instead we assumed, without justification, that \texttt{open door} returns Success \emph{immediately whenever} the door is open. That is, we read the BT as if it were the reactive selection BT \texttt{(doorOpen ? open door) $\to$ passThrough door}. Nowhere was it prescribed that \texttt{open door} has this behavior, and it actually seems unlikely---\texttt{open door} might well, for instance, take several ticks to move the robot to scan the state of the door, before establishing that the door is open. In that case we would still have the progress problem. Or does the robot only return to \texttt{open door} when it `notices' that the door is closed? Worse, how should we interpret the subtree \texttt{open door $\to$ passThrough door $\to$ close door}? It seems like, on the tick after we close the door, we tick \texttt{open door} which, observing that the door is currently closed, reopens it, sending us into a loop.

This fallacy arises from assuming that every BT is implicitly equivalent to one in reactive selection form. That is, we each Action (and consequently each subtree) exists to solve a specific task with specific pre- and post-conditions, and upon ticking, does the following: check postconditions, if satisfied return Success; check preconditions, if violated return Failure; otherwise do task. This interpretation avoids the progress problem, but often doesn't make sense. Many behaviors don't have pre- and post-conditions, or they can't be checked in a single tick. The result is ambiguous interpretations of the BT, which goes against one of the key motivations for BTs. In Sections~\ref{sec: when to use react} we explain why, in many cases like this one, an explicitly non-reactive design is better.

Note that robotics and game AI differ here. In game AI, the assumption of pre- and post-conditions is sometimes reasonable; in the example above, the agent can possibly query the `world' to determine if the door is open within the span of a single tick.

To avoid the progress problem, two well-studied BT design techniques exist to explicitly transform BTs into reactive selection form. These are: \textbf{explicit success conditions}, which involves writing the pre- and post-conditions as explicit Conditions in the tree, and \textbf{backchaining} \cite{btbook,ogren2020convergence}. Backchaining involves replacing an action \texttt{action} with the subtree \texttt{postcondition ? (precondition $\to$ action)}. Its semantics are ``check postconditions, if satisfied return Success; check preconditions, if violated return Failure; otherwise do task". Given a sequence of actions \texttt{action1, action2, ...} where the postcondition of an action's successor satisfy its preconditions, we can `chain' them together as \texttt{postcondition1 ? ((precondition1 ? (((precondition2 ? (...)) $\to$ action2) $\to$ action1)}. Note that the resultant BTs are in reactive selection form, avoiding the progress problem. Applying this solution to the subtree \texttt{open door $\to$ passThrough door $\to$ close door} gives us the more complicated tree\footnote{Note that we have to assume the overall subtree has the goal of being \texttt{insideRoom \& doorClosed} to avoid the problem of reopening the door after it is closed.}:
\begin{align} \label{subtree translation}
\begin{split}
    \tt (insideRoom \And doorClosed) &\ \fb \\
    \tt ((insideRoom \fb ((doorClosed \fb open~door) & \to \\ \tt passThrough~door)) & \to \tt close~door) \to doTask
    \end{split}
\end{align}
This BT can be written more clearly as 
\begin{minted}{haskell}
passDoor = rSelect (insideRoom & doorClosed) doTask
            rSelect insideRoom (close door)
             rSelect doorClosed (open door) (passThrough door)
\end{minted}

Of course, we have replaced our original BT with a \emph{different tree}, and we needed to make a number of additional assumptions about the tree to get it in this form. In other words, the original non-reactive selection form BT was ambiguous.

\subsection{When is reactiveness appropriate?} \label{sec: when to use react}

In the previous section we established that, because of the progress problem, we can only unambiguously interpret reactive BTs when they are in reactive selection form, where they reduce to reactive Decision Trees. We think that writing them explicitly in this form is clearer than using nested \sequence{} and \fallback{}.

If we are given a BT which is not in reactive selection form, and we want to express it unambiguously, we need to translate it into either non-reactive operations, or reactive selection form. Often, both are possible: in this section we discuss the tradeoffs involved in choosing one or the other.
We can summarise this section as follows: \textbf{we should use reactiveness only when we explicitly need some behaviors to \emph{interrupt} other behaviors} and in that case we should make very clear the circumstances under which interruption takes place.

We start with an example where interruptions, and so reactiveness, is absolutely necessary. Our example involves handling prioritised tasks, like the \texttt{(battery>10\% ? recharge) $\to$ task} subtree in Fig~\ref{fig:btbooktree}, which in reactive selection form could be written as 
\begin{minted}{haskell}
bt = rSelect (battery > 10) task recharge
\end{minted}
Here, the reactiveness is giving us a modular and readable behavior which immediately interrupts \texttt{task} when the battery is low, without the designer of \texttt{task} needing to explicitly check for it. It is easy for us to reason that the robot will not run out of charge, because we are certain to respond to the battery level quickly, regardless of what \texttt{task} is\footnote{There is a problem here: the condition \texttt{battery>10\%} interrupts \emph{both} \texttt{task} and \texttt{recharge}, when in reality we want to \texttt{recharge} fully without being interrupted. We discuss this problem and its solution in Section~\ref{sec: chattering}.}. Another example where reactiveness is needed is responding to user input, where we might need to interrupt the running task to respond.

Other times, a non-reactive behavior makes more sense. For instance, if we are following waypoints: \texttt{ moveTo point1 $\to^*$ moveTo point2}. Most programming happens with a sequence of instructions of this form, where the task is simply to do one thing after another. No interruptions are necessary here.

In theory, we never need an explicit \sequence{} or \fallback{} with memory, because a construction exists~\cite{btbook} to replace these with reactive \sequence{} or \fallback{} nodes, and the resultant BTs are in reactive selection form. Specifically, we replace \texttt{action1 $\to^*$ action2} with \texttt{(action1done ? action1) $\to$ action2} (equivalently, \hask{rSelect action1done action2 action1}). In the example above, we might get \texttt{(reachedPoint1 ? moveTo point1) $\to$ moveTo point2}. This isn't a very useful translation, however. We have replaced a simple and \emph{clearly non-reactive} behavior with a more complicated tree which \emph{looks like} it defines a reactive behavior, but doesn't. As we set out in the introduction, our goal is to present non-reactive and reactive behavior in a readable and well-delineated way---this construction doesn't do that.

More concretely, suppose we want to apply this transformation to the BT \texttt{circle $\to^*$ goto location} which drives the robot in a circle, then moves it to \texttt{location}. We face the problem that the \texttt{circle} behavior ends with the robot in the same place as it started, so there is no obvious `external' condition to indicate that it has completed. We could construct an `artificial' condition, like \texttt{circleDone} to check if circling has finished. We obtain \texttt{(circleDone ? circle) $\to$ goto location}. Note the tautology: it runs \texttt{circle} if and only if \texttt{circle} is already running. This is a circular\footnote{Pun intended.} argument, and it is ambiguous, because we haven't explained how the value \texttt{circleDone} is modified\footnote{The question of how BTs modify any variables in the topic of Section~\ref{sec: data}.}. Consequently, it is highly error-prone---for instance, suppose this subtree is interrupted: does \texttt{circleDone} get reset or not? Our original non-reactive BT was much clearer.

This is an extreme example. Often, the question of whether to express a behavior using non-reactive operations, reactive selection or a combination of both is less clear-cut. For instance, in the previous section we examined the subtree \texttt{open door $\to$ passThrough door $\to$ close door}. Using explicit success conditions, we translation this to the BT in~\eqref{subtree translation}. How does this compare to the much simpler non-reactive form \texttt{open frontDoor $\to^*$ passThrough frontDoor $\to^*$ close frontDoor}? While more complicated, a reactive behavior has two potential advantages: we could \textbf{skip unneeded actions}, e.g. if the door is already open we do not attempt to open it, and we \textbf{automatically handle failure of preconditions}, so if the door closes while we are passing through it we immediately switch to \texttt{open door}. However, we do not get these properties for free, and designing them is more subtle than it first appears. In the next section, we discuss the trade-offs between reactive and non-reactive solutions, and how to design them transparently.

\subsection{Pre-empting Success and Failure} \label{sec: preempting}

One benefit of reactiveness, in the BT sense, is that we can skip actions that are not necessary ~\cite{btbook}. For instance, in the above example, if the robot is in the room we can skip \texttt{open door} and \texttt{passThrough door}. Likewise, we can automatically recover from failures of preconditions, so, for instance, if the door closes while we are passing through it we immediately switch to \texttt{open door}.

However, these benefits are not free. First, they only apply to behaviors whose Success and Failure can be represented by \emph{explicit success conditions}. As we discussed in the \texttt{circle} example above, some behaviors do not have meaningful external conditions representing their Success or Failure. Even in our door example, these conditions can be questionable. Consider
\begin{minted}{haskell}
bt = rSelect insideRoom (close door) (passThrough door)
\end{minted}
This pre-emptively skips or interrupts \texttt{passThrough door} if its post-condition \texttt{insideRoom} is true. A benefit of this approach is that if we achieve entry into the room unexpectedly, for instance because a human operator moved the robot while \texttt{passThrough door} is running, we can interrupt it and move immediately to \texttt{close door}. In some applications, this may be useful, but here it is probably too optimistic---do we need to account for this unexpected way to achieve success? Depending on its implementation, \texttt{passThrough door} may return Success in this case anyway. This comes at a cost too. We implicitly assumed that \texttt{passThrough door} was successful if and only if the condition \texttt{insideRoom} was true. While \texttt{passThrough door} should result in the robot being in the room, a good implementation would actually bring us some distance into the room, rather than pre-emptively interrupting \texttt{passThrough door} the moment we cross the threshold. In this case, it is probably simpler and more modular to use the non-reactive \texttt{passThrough door $\to^*$ close door}, allowing \texttt{passThrough door} to determine the distance it brings us into the room before succeeding, rather than interrupting it. Contrast this with a case like
\begin{minted}{haskell}
bt = rSelect closeToWall turn moveForward
\end{minted}
Here we need the interruption: \texttt{moveForward} doesn't determine for itself when it should stop.

If checking \texttt{insideRoom} is complex, we could still skip \texttt{passThrough door} using a normal, non-reactive \texttt{if}-statement and non-reactive sequencing:
\begin{minted}{haskell}
bt = do if not insideRoom then passThrough door else return ();
        close door
\end{minted}
Here \hask{return ()} is a null action in Haskell\footnote{Actually, the Haskell function \hask{unless} provides a neater version of this: \hask{do unless insideRoom (passThrough door); close door}}. This version has no interruptions; it checks, once, if the robot is already inside the room and if not runs \texttt{passThrough door}, then runs \texttt{close door}.

A similar argument applies for pre-empting failure of pre-conditions. For instance, we assumed that \texttt{doorOpen} is a precondition of \texttt{passThrough door} and post-condition of \texttt{open door}. But this variable is not binary---\emph{how far} open does the door need to be? Does this mean `fully' open? It might not need to be; \texttt{passThrough} probably has a tolerance threshold to have its pre-conditions satisfied. We could replace it in~\eqref{subtree translation} with a more-specific Condition \texttt{preconditionPassThrough}, so the BT stops \texttt{open door} the moment this threshold is reached\footnote{The door is always left in a marginal position---we must continually monitor it while performing \texttt{passThrough door}, and if it drifts to even partially closed we return to \texttt{open door}. We discuss the resultant `chattering' problem in Section~\ref{sec: chattering}.}, but it probably makes more sense for \texttt{open door} to open the door fully before succeeding. Realistically, we probably can't express \texttt{preconditionPassThrough} as a condition in advance anyway, and \emph{more importantly, we don't want to, because this undermines modularity}.

Trying to pre-empt Success and Failure required us to unpack the pre- and post-conditions of \texttt{passThrough door}. However, the point of using a common interface between behaviors is \emph{precisely so that} the designer does not need to know exactly how a behavior succeeds or fails. We are trying to use, effectively, huge nested \texttt{if}-statements to pre-emptively catch all the failure cases, and this considered bad practice in programming. The better solution for catching unexpected failure cases is to use Exceptions. In BTs, we should \emph{allow the behavior to fail, and use (non-reactive) \fallback{} to recover}. Note that using the non-reactive \fallback{} in this context is actually still `reactive' in the intuitive sense---it responds immediately to Failure with an appropriate change in behavior.

Let's look at another example from Figure~\ref{fig:btbooktree}. Suppose the action \texttt{open frontDoor} failed, perhaps because the door is locked. We fall back to \texttt{open backDoor}. As written, we run into the progress problem because we continue to tick \texttt{open frontDoor}. From a conceptual perspective, there is no reasonable way to turn this into a reactive selection---it only makes sense as a non-reactive behavior. While the robot is recovering from the failure by attempting the back door, there is no condition we could check which would tell us whether the front door \emph{would} now open if we attempted it again.

Trying to use a reactive selection to recover from Failure requires pre-emptively knowing why a behavior failed. We think that almost all of the time, the non-reactive \fallback{} ($?^*$) is a better solution, and it should be used in the same way we would use a \hask{catch} statement to handle Exceptions in a standard progamming language. In our behavior language, the \fallback{} operation \hask{?} always corresponds to non-reactive fallback, and it is an infix synonym for Haskell's built-in \hask{catch} function.

    

\subsection{Chattering and one-sided reactiveness} \label{sec: chattering}

Let's go back to the first example we used from Figure~\ref{fig:btbooktree}, which is
\begin{minted}{haskell}
    bt = rSelect (battery > 10) task recharge
\end{minted}
As we discussed in the previous section, handling prioritised tasks (in this case by ensuring we recharge whenever necessary) is one of the key use cases where reactiveness is needed. Specifically, we need to constantly check the battery level, and interrupt the \texttt{task} when that level is too low.

However, reactive selections, like the \texttt{if}-statements they are based on, are symmetric with respect to the condition. So even though we interrupt \texttt{task} when the battery drops less than 10\%, we symmetrically interrupt \texttt{recharge} when the level rises above 10\%. This leads to chattering: the battery increases above 10\%, so we resume \texttt{task}, then decreases below 10\%, so we recharge, then increases above 10\%, so we resume the task, and so on. We cannot avoid this behavior using only the reactive BT operations. But a non-reactive behavior also doesn't work---we want \emph{one-directional} reactiveness, so the battery level interrupts \texttt{task} but cannot interrupt \texttt{recharge}. 

We call this situation \emph{monitoring}, where we have prioritised tasks which can interrupt each other only in the order of priority. This is the idea behind \emph{Teleo-reactive programs} \cite{nilsson1993teleo}. Precisely, we want the following semantics.
\begin{defn}[Monitor]
    \hask{monitor} is a control flow node which takes three arguments: a condition \texttt{test}, and two behaviors \texttt{recovery} and \texttt{task}. The behavior \texttt{monitor test recovery task} works as follows.
    \begin{enumerate}
        \item If the condition \texttt{test} is false, we run \texttt{task}. While running \texttt{task} we monitor the value of \texttt{test}, and if true we immediately switch to \texttt{recovery}.
        \item If the condition \texttt{test} is true, we run \texttt{recovery}. We always run \texttt{recovery} to completion, regardless of the value of \texttt{test}, and when complete we again run \texttt{monitor test recovery task}.
        \item If either \texttt{recovery} or \texttt{task} fail then \hask{monitor test recovery task} fails.
        \item \hask{monitor test recovery task}\footnote{In Haskell, brackets are only needed to resolve ambiguity, and not for applying a function to its arguments. In a C-like language we might write this as \texttt{monitor(test, recovery, task)}.} succeeds if \texttt{task} completes successfully.
    \end{enumerate}
\end{defn}
In the context of our battery example, this would be \texttt{monitor batteryLow recharge task}. While running \texttt{task} we monitor the battery level, and whenever it becomes `low' we switch to \texttt{recharge}, which is run \emph{to completion}, and then the whole tree is resumed. There is no chattering, and we can recharge as many times as necessary while performing the task. The subtree completes if we successfully complete \texttt{task}.

We can't implement \hask{monitor} with the existing BT operations, without using additional variables. Even if we could define operations recursively (generally not possible in BT libraries) the closest analogue is:
\begin{minted}{haskell}
    monitor test recovery task = (test ?* task) ->*
                                 recovery ->*
                                 monitor test recovery task
\end{minted}
This solution nearly works, but doesn't satisfy our fourth criterion for \hask{monitor}. Instead, if \texttt{task} succeeds we run \texttt{recovery} rather than succeeding. In fact, this definition never completes successfully. 

Safety examples like this battery one are ubiquitous uses of BTs. Additionally, \hask{monitor} provides a neat solution to several of our previously-discussed examples. For instance, recall
\begin{minted}{haskell}
bt = rSelect insideRoom (close door) (passThrough door)
\end{minted}
We pointed out that \texttt{passThrough door} should not interrupted the moment the robot crosses the threshold, and should be allowed to move the robot further inside the room before succeeding. Suppose that we also needed to respond to a human repeatedly moving the robot out of the room. A neat solution achieving both criteria is
\begin{minted}{haskell}
    bt = monitor insideRoom (passThrough door) (close door)
\end{minted}
Here, the \texttt{test} is \texttt{insideRoom}, \texttt{task} is \texttt{close door} and \texttt{recovery} is \texttt{passThrough door}. Hence, if a human moves the robot outside the room while running \texttt{close door}, we immediately run \texttt{passThrough door} to completion, then resume closing the door.

How do existing implementations avoid chattering in this very common example? A standard solution~\cite{btbook} involves an auxiliary variable, as follows: \texttt{((batteryHigh \& not recharging) ? recharge) $\to$ task}. Similarly to previous auxiliary-variable solutions, we have tautological semantics: we perform \texttt{recharge} if either the battery level is low or \emph{we are recharging}. This presented approach is a mixture of specification and implementation, describing a programming technique used to \emph{implement} the behavior described by \hask{monitor}---specifically, using an auxiliary variable---but these semantics are not present here, because we don't know how the variable \texttt{recharging} is modified. Correctly maintaining the value of this variable is non-trivial, especially in the presence of concurrency. We avoid ambiguity by specifying clearly what we desire from \hask{monitor} as an operation. Any particular implementation of \hask{monitor} can choose how to achieve this specification. We discuss our implementation in Section~\ref{sec: specification}.


Considering the importance of one-sided interruptions, and the complexity of correctly implementing \hask{monitor}, we think it should be a core part of the toolkit provided by a behavior language for managing interruptions. In fact, along with \rselect{} (`two-sided interruptions'), it is one of the two operations we provide for building reactive behavior. See Section~\ref{sec: specification}.

\section{Data} \label{sec: data}

Programming languages have two main roles: structuring instructions, and organising data. Behavior Trees, with their roots in robotics and AI, have historically focused on the former. This makes sense; many simple behaviors are constructed simply from a sequence of instructions, and the agent does not need to pass data around.

However, as robotic behaviors become more complex, manipulating data becomes just as important as in classical programming languages. For instance, many actions take data as an argument: in the previous section, the behaviors \texttt{open}, \texttt{passThrough} and \texttt{close} take a \texttt{door} object as an argument; \texttt{moveTo} takes a target \texttt{location}; and so on. Similarly, many actions also \emph{produce} data: for instance, we might use an action \texttt{findDoors} to identify which doors are available; \texttt{locateItem} to find an item; or \texttt{batteryLevel} action to obtain the battery level as a number. If our subsequent actions or decision-making depends on the output of our current action, then we must have a mechanism for passing data within a BT.

This point has been recognised for a long time in the BT literature, and modern BT libraries have a number of approaches for handling data~\cite{colledanchise2021implementation}. The classical approach is to use a Blackboard \cite{btbook}. Essentially, a Blackboard is a location where variables can be freely written to and retrieved. That is, all variables are stored globally with no restrictions on access\footnote{The global variable aspect of Blackboards is known to be a problem, with several suggested modifications to mitigate this \cite{colledanchise2021implementation,isla2015handling,iovino2022survey,ghzouli2020behavior}. However, generally, these methods are not particularly modular (cf. \cite{btcpp}).}. Unfortunately, blackboards can be quite opaque, and they are not modular. In programming, we have known for a long time that global variables are not a good idea, because they break the modularity of structured programs, leading to cryptic bugs and generally difficult-to-read programs. 

Because modularity is the critical property which separates BTs from FSMs, we want to pass data in a way which is as modular as possible. 
We choose a simple solution: behaviors are simply \emph{asynchronous functions}, which can both take arguments and return values. When we want one behavior to have access to a value from another, we use \emph{function composition}. In fact, because we view modularity as such a critical property in BTs, we take quite a strict approach to passing values: we \emph{only} pass values through composition. This is the approach taken by Haskell.

BTs combine behaviors using a small collection of operations. We want to integrate these operations with function composition. The first observation we need is that actually only one operation needs to be modified to handle data: (non-reactive) sequence. This is because this is the only operation which waits for its first argument to complete, successfully, before running its second argument. By contrast, (non-reactive) fallback runs its second argument when the first \emph{fails}. In programming, a failed action returns an Exception rather than a value of its declared type. The reactive operations \rselect{} and \hask{monitor} interrupt their arguments before they might return. Hence, all we need to establish is how non-reactive sequence combines with data. Specifically, we want \texttt{action1 $\to^*$ action2} to run \texttt{action1}, pass its return value to \texttt{action2}, then run \texttt{action2} with that value as its argument, similar to the composition \mintinline{python}{action2(action1())}.

Let's look at an example. Suppose we want to construct a behavior which explores a room to find a box, then moves to the box. We have available the actions \texttt{findBox} and \texttt{moveTo box}. \texttt{moveTo} takes a \texttt{box} argument, while \texttt{findBox} returns a \texttt{box} object. We want \texttt{findBox $\to^*$ moveTo} to have the semantics of 
\mintinline{python}{moveTo(findBox())}. 

However, there is one important difference between \texttt{findBox $\to^*$ moveTo} and \mintinline{python}{moveTo(findBox())}. The BT operations work at compile-time, fusing the actions \texttt{findBox} and \texttt{moveTo} into a single action, which can be passed around and fused with other actions before the action is actually run. It turns out, very neatly, that this is exactly how actions (functions which perform I/O), are composed in Haskell~\cite{marlow2010haskell}, and it is for this reason that we choose Haskell to implement our behavior library. In Haskell, the \texttt{bind} function, written infix as \hask{>>=}, has exactly the semantics we wanted from non-reactive sequence: \hask{bt = findBox >>= moveTo} returns a single action, which, when it is eventually executed, runs \texttt{findBox}, and passes its result to \texttt{moveTo}. If we want to give the variable a name, we can use a anonymous (lambda) function:
\begin{minted}{haskell}
    bt = findBox >>= (\ box -> moveTo box)
\end{minted}
Haskell provides an imperative-like syntax for this construction, called \hask{do}-notation. The above is exactly equivalent to:
\begin{minted}{haskell}
    bt = do box <- findBox
            moveTo box
\end{minted}
While this looks like two statements, one assigning the output of \texttt{findBox} to a variable \texttt{box} and the next calling \texttt{moveTo}, this is only syntactic sugar for the previous example. That is, neither action is executed at this line, they are only fused into a single action called \hask{bt}.

To ensure safe composition of behaviors, BTs declare and conform to an interface, consisting of the values Success, Failure and Running. In the same way, we want our behavior type to declare an interface. Again, this has a well-established programming solution: behaviors, as asynchronous functions, have a \textbf{return type}. So, for instance, a behavior which returns an integer would have a type of \hask{Behavior Int}. The \behavior{} indicates this function performs some action in the world, eventually returning with an integer result\footnote{In our implementation, \behavior{} is a synonym for Haskell's built-in type \hask{IO}.}. In general, we give behaviors a type of \hask{Behavior a}, where \texttt{a} is a type parameter.

Now that behaviors are explicitly asynchronous and have a return type, the original Success, Failure and Running values are actually unnecessary. These values are tools to make the tick mechanism work, but our reactive operations run concurrently and do not use the tick mechanism. We can tell if a behavior is running simply by observing that it has not returned control to the caller, and we can tell if it has failed or succeeded by determining whether it raised an Exception or returned a value of its declared type.

What types do the operations have in our box-finding example? Using Haskell's notation, the actions \texttt{moveTo} and \texttt{findBox} would have types:
\begin{minted}{haskell}
    findBox :: Behavior Box
    moveTo :: Box -> Behavior ()
\end{minted}
Here \hask{Box} represents a box object, and the special type \hask{()} represents a null value, because \texttt{moveTo} doesn't return data. That is, \texttt{findBox} is an action which returns a \hask{Box}, and \texttt{moveTo} takes a \hask{Box} object and returns an action which doesn't return any value. We joined them using the operation \hask{>>=}, which has polymorphic type:
\begin{minted}{haskell}
(>>=) :: Behavior a -> (a -> Behavior b) -> Behavior b
\end{minted}
This function takes two \behaviors: the first returns a value of type \texttt{a}, and the second takes an argument of type \texttt{a} and returns a \hask{Behavior} which returns a \texttt{b}. \hask{>>=} composes these \behaviors{} together, resulting in an overall \behavior{} which returns a value of type \texttt{b}. In our example, \hask{>>=} has the concrete type:
\begin{minted}{haskell}
(>>=) :: Behavior Box -> (Box -> Behavior ()) -> Behavior ()
\end{minted}
Then \hask{bt = findBox >>= moveTo} has type \hask{bt :: Behavior ()}. Using \hask{>>=} for non-reactive sequencing gives us a simple and elegant solution for passing data between behaviors in a modular way. In our library, the sequence with memory operation $\to^*$ is exactly Haskell's bind operation \hask{>>=}, and we compose behaviors just as Haskell composes actions.

\subsection{Booleans and branching} \label{sec: booleans}

The existence of return types instead of the flags Success, Failure and Running allows us to avoid a problem which is present in many classical BTs. Generally, the values Success and Failure are used implicitly as boolean values, especially in Condition nodes. Again, this is avoided in programming, and there is a good reason why: we can't distinguish between Failure, indicating our behavior did not work, and False, indicating the Condition is not true. Conditions, like any other robotic action, can fail for many reasons, and we need to identify when this has happened.

The following example demonstrates the severity of the problem. Consider a Condition \texttt{dangerNearby}, which detects some kind of threat to the robot. We build this into a (reactive selection form) BT: \texttt{(dangerNearby ? doTask) $\to$ hide}. However this BT can't distinguish between \texttt{dangerNearby} \emph{failing}, possibly because some sensors are broken, and it \emph{successfully returning} with the value False, indicating there is no danger nearby. In both cases we run \texttt{doTask}.

Luckily, we can now avoid this problem: a Condition is a behavior whose type is \hask{Behavior Bool}, that is, it returns a boolean. Recalling our discussion in Section~\ref{sec: chattering}, we would probably write this BT using \hask{monitor}:
\begin{minted}{haskell}
bt = monitor dangerNearby hide doTask
\end{minted}
Now, if \texttt{dangerNearby} fails, the overall behavior fails, and we can address the problem appropriately using a \fallback{}.

\section{Specification for the Behavior Language} \label{sec: specification}

Over the previous sections, we discussed how to distill the concepts of reactiveness and modularity into a programming language for behavior. In this section we lay out the requirements of such a language. In particular, we explain the concepts in our Haskell implementation. While we believe Haskell is a particularly neat language for implementing these operations, the key contribution here is the ideas and operations, which could be implemented in any language.

Recall that we needed a behavior language to be first and foremost a language, equipped with all the control structures and tools that we expect of any language, including the ability to name functions and variables, to write loops and to branch. From here, we will focus on the behavior-specific operations like reactiveness, sequencing and fallback, which we have discussed throughout this paper.

\begin{itemize}
    \item \textbf{Behaviors}
    
    The first piece of our Behavior language is the \behavior{} type, which encapsulates a robotic action which returns a type. \behaviors{} may read from input, produce output and execute for an arbitrary amount of time. \hask{Behavior a} is a type constructor which takes a single type argument \texttt{a}, representing the return type of the behavior. A behavior which returns a floating-point number, for instance, has type \hask{Behavior Float}. \behavior{} is a type synonym for the built-in Haskell type \hask{IO}. A behavior with arguments has the form of a function returning a behavior, so for instance a behavior \texttt{moveTo} might have type 
    \hask{moveTo :: Location -> Behavior ()}, indicating that it takes a \texttt{Location} and returns a \behavior{} with the null type \hask{()}.
    
    \item \textbf{Sequencing Behaviors with data}
    
    We the simplest way of joining \behaviors{} is the (non-reactive) sequence operator $\to^*$. For this, we use the Haskell function \texttt{bind}, written infix as \hask{>>=}. This glues two IO actions into a single one, passing the return value from the first to the second.
    \begin{minted}{haskell}
    (>>=) :: Behavior a -> (a -> Behavior b) -> Behavior b
    (>>=) = bind
    \end{minted}
    Specifically, running \hask{bt = bhvr1 >>= bhvr2} runs \hask{bhvr1 :: Behavior a} to completion, then hands the result (of type \hask{a}) to \hask{bhvr2 :: a -> Behavior b}, runs it and returns the result of type \hask{b}. This is a fundamental construction in Haskell, and so a special `imperative-like' syntax is provided for it, called \texttt{do-notation}. \hask{bhvr1 >>= bhvr2} can be written: 
    \begin{minted}{haskell}
    do result <- bhvr1
       bhvr2 result
    \end{minted}
    This desugars to
    \begin{minted}{haskell}
    bhvr1 >>= (\ result -> bhvr2 result)
    \end{minted}
    
    \item \textbf{Sequencing Behaviors without data}
    
    Behaviors may not always return values, or sometimes we wish to explicitly ignore those values. Haskell provides a special notation for this: \hask{>>}. Specifically, \hask{bt = bhvr1 >> bhvr2} is equivalent to \hask{bt = bhvr1 >>= \_ -> bhvr2}, which binds \texttt{bhvr1} with a function which throws away its argument and runs \hask{bhvr2}. In \hask{do}-notation:
    \begin{minted}{haskell}
    do bhvr1
       bhvr2
    \end{minted}
    
    \item \textbf{Fallback/Catch}
    
    Per our earlier discussion, Fallback is specifically designed to catch errors and is not reactive. We provide both a function form and and infix operator form.
    \begin{minted}{haskell}
    fallback :: Behavior a -> Behavior a -> Behavior a
    fallback bhvr1 bhvr2 = catch bhvr1 
        (\(e :: BehaviorFailed) -> bhvr2)
    (?) = fallback
    \end{minted}
    Running \hask{bt = bhvr1 ? bhvr2} runs \hask{bhvr1}, and if a \hask{BehaviorFailed} exception is raised it runs \hask{bhvr2}. The returned value (of type \hask{a}) is either the result of \hask{bhvr1} or \hask{bhvr2}, hence they must be of the same type.
    
    \hask{fallback} is just a synonym for \hask{catch} which only catches Exceptions whose type is an instance of \hask{BehaviorFailed}. One may want to filter explicitly for certain subtypes of \hask{BehaviorFailed}, for which an explicit \hask{catch} can be used.
    
    \item \textbf{Parallelism}
    
    Running two Behaviors in parallel is a simple way of combining agent behaviors, and a well-established component of Behavior Trees. We provide two binary parallel operations: \hask{parallel}, which races its arguments until the first completes, then cancels the other, and \hask{both}, which waits for both arguments to complete.
    
    \begin{minted}{haskell}
    parallel :: Behavior a -> Behavior b -> Behavior ()
    (|||) = parallel
    \end{minted}
    \hask{parallel} runs two actions in parallel and completes when the first of them completes, throwing away the result. We provide \hask{|||} as an infix syntax for parallel.
    
    \begin{minted}{haskell}
    both :: Behavior a -> Behavior b -> Behavior ()
    \end{minted}
    \hask{both} runs two actions in parallel and waits for both to complete, throwing away the result.
    
    \item \textbf{Reactive operations}
    
    Reactiveness is a core concept for Behavior Trees, as we discussed in Section~\ref{sec: reactiveness}. We provide two operations for constructing reactive behavior, \rselect{} and \hask{monitor}.
    
    The operation is \rselect{}, which is essentially a reactive \texttt{if}-statement.

    \begin{minted}{haskell}
rSelect :: Behavior Bool -> Behavior a -> Behavior a -> Behavior a
    \end{minted}
    \rselect{} takes three arguments: \texttt{test}, a Condition (a \hask{Behavior Bool}) to be run repeatedly for its Boolean output, and two behaviors \texttt{left} and \texttt{right} to select between. \texttt{left} and \texttt{right} do not need to be able to run in parallel with each other, as they will never be simultaneously running. However, they must have the same type, because \rselect{} completes when either \texttt{left} or \texttt{right} runs to completion and returns the value from the completed behavior.
    
    Note that \rselect{} treats its arguments \texttt{left} and \texttt{right} symmetrically---both can equally be interrupted by \texttt{test}. When we want \emph{one-sided} interruptions (Section~\ref{sec: chattering}), we use \hask{monitor}.
    
    \begin{minted}{haskell}
monitor :: Behavior Bool -> Behavior () -> Behavior a -> Behavior a
    \end{minted}
    Like \rselect{}, \hask{monitor test recovery task} repeatedly runs \texttt{test}, and when false runs \texttt{task} while checking \texttt{test} in parallel. However, whenever \texttt{test} becomes true, we halt \texttt{task} and run \texttt{recovery} \emph{to completion} without monitoring \texttt{test}. When \texttt{recovery} completes, \hask{monitor} resumes. Note the type signature: \texttt{recovery} is required to return a null value \hask{()} because its result cannot be used---after completing \texttt{recovery}, we resume monitoring. \hask{monitor} completes when \texttt{task} completes.
    
    
    
\end{itemize}


\section{Example} \label{sec: example}

In this section we demonstrate how we can use our library to rewrite the BT in Figure~\ref{fig:btbooktree} as executable Haskell code. First, we outline the available behaviors and objects:
\begin{minted}{haskell}
    batteryLow :: Behavior Bool
    recharge :: Behavior ()
    open :: Door -> Behavior ()
    passThrough :: Door -> Behavior ()
    close :: Door -> Behavior ()
    smash :: Door -> Behavior ()
    doTask :: Behavior ()
    isOpen :: Door -> Behavior Bool
    frontDoor :: Door
    backDoor :: Door
\end{minted}

Now, the most important step: we must (precisely!) describe what we want from our robot. Recalling our discussion in Section~\ref{sec: ticking}, we choose the following interpretation.

First, the robot must never run out of battery; whenever the battery drops below 10\% we should stop what we are doing and recharge fully, then resume. Second, to complete the task, we must enter a room, which is accessible via two doors, \texttt{frontDoor} and \texttt{backDoor}. We will first attempt the front door. We open it and pass through, unless it is already open, in which case we will skip that step. If we pass through it successfully, we close the door behind us and do the task. If we fail to enter via the front door, for any reason, we attempt the back door, opening it then passing through. If opening the door fails, then we attempt to smash it. If this fails, or passing through fails, the overall BT should fail. Otherwise, we perform the task.

We could implement this as:

\begin{minted}{haskell}
bt = monitor batteryLow recharge
        (( (if (isOpen frontDoor) then (open frontDoor) else return ()
          >> passThrough frontDoor
          >> close frontDoor)
          ? ((open backDoor ? smash backDoor)
            >> passThrough backDoor))
        >> doTask)
\end{minted}
Per Section~\ref{sec: chattering}, we have used a \hask{monitor} to ensure that we recharge whenever necessary. Also, from our discussion in~\ref{sec: when to use react} we used an \texttt{if}-statement to skip the action and interpreted all the other steps using non-reactive operations.

This version uses only the most basic operations. However, because our language is embedded in Haskell, we have the full power of a programming language at our disposal! For instance, making use of some built-in Haskell constructs (\hask{unless}, \texttt{\$}\footnote{The \texttt{\$} is the function application operator. It is used, as we did in this example, to remove a layer of brackets. So, for instance \hask{f (x + y)} can be rewritten \hask{f $ x + y}, which helps prevent many layers of nested brackets in functional code.}, and \hask{do}-notation), we can make this neater:
\begin{minted}{haskell}
bt = monitor batteryLow recharge $
        (do unless (isOpen frontDoor) (open frontDoor)
            passThrough frontDoor
            close frontDoor)
        ? (do (open backDoor ? smash backDoor)
              passThrough backDoor)
        >> doTask
\end{minted}
Like Figure~\ref{fig:btbooktree}, this describes a complex sequence of actions in a readable way. Now, however, it is precise, executable code.

We can now extend this example in ways which would have been challenging in the classical BT setup. First, we can use return values to build \texttt{batteryLow} from a \hask{batteryLevel :: Behavior Int} behavior, which returns the remaining battery as an integer from 0 to 100. We could use:
\begin{minted}{haskell}
    batteryLow = (<10) <$> batteryLevel
\end{minted}
Here we used a Haskell function \texttt{<\$>}, which is function application over actions. Specifically, we applied the integer output of the \behavior{} \texttt{batteryLevel} into the (non-\behavior) \hask{(<10)}, which just returns whether the value is less than 10. Now, if we wanted to change the battery threshold, we can just modify the value 10.

Our code currently has some repetition for the cases of \texttt{frontDoor} and \texttt{backDoor}. We can avoid this by mapping our behavior over a list of \hask{Door}s. Suppose we have a \behavior{} called \hask{findDoors :: [Door]} which returns a list of available doors. For each door, we try to open it, pass through, and close it, and if we fail we move on to the next door. If we do succeed, we close the door behind us. To do this, we first need a version of \hask{fallback} which can apply a \behavior{} to a list of inputs, falling back to the next input\footnote{We can give this function the same name `\hask{fallback}', because Haskell can distinguish the different versions on the basis of their type signatures.}. This is easy to implement:
\begin{minted}{haskell}
fallback (x:xs) bhvr = bhvr x ? fallback xs bhvr
\end{minted}
Using this, we can write the overall BT as:

\begin{minted}{haskell}
openPassClose door = open door >> passThrough door >> close door
enterRoom = do doors <- findDoors
               fallback doors openPassClose
monitorBattery = monitor ((<10) <$> batteryLevel) recharge

bt = monitorBattery (enterRoom >> doTask)
\end{minted}
By extending the BT concepts into a full programming language, we can implement complex BTs like Figure~\ref{fig:btbooktree} in just a few lines of simple and readable code. Additionally, it is equally easy to extend this model using new functions without requiring building new `control flow nodes' or `Decorator nodes'. For instance, a common use of a Decorator node in BTs is to implement behaviors like `attempt $n$ times'. In this example, we might want to try to enter each door three times before moving on to the next door. This would look something like:
\begin{minted}{haskell}
openPassClose door = open door >> passThrough door >> close door
enterRoom = do doors <- findDoors
               fallback doors (attempt 3 openPassClose)
monitorBattery = monitor ((<10) <$> batteryLevel) recharge

bt = monitorBattery (enterRoom >> doTask)
\end{minted}
The \hask{attempt} behavior can be implemented with a single line of code:
\begin{minted}{haskell}
attempt n task = if n == 0 then throw Failed else task ? attempt (n-1) task
\end{minted}
\section{Conclusions}

In this paper we discussed the challenges of extending BTs to full programming language, and presented some modular solutions. This is summarised in our overall language specification, and accompanying implementation in Haskell. We hope that our findings will provide insight for the process of developing a modern approach to programming autonomous behavior.

There is an immediate direction for future work: constructing a full implementation capable of performing non-trivial tasks on real hardware, and ideally interfacing with pre-existing tools like ROS~\cite{macenski2022robot}. This would take significant effort, but could be of great value, given the interest in BTs. 
\bibliographystyle{plain}
\bibliography{refs}

\end{document}